\documentclass[10pt]{article}
\usepackage{latexsym}
\usepackage{amsmath}
\usepackage{indentfirst}
\usepackage{cite} 

\long\def\ca#1\cb{} 

\newcommand{\becs}{\begin{cases}}
\newcommand{\bem}{\begin{matrix}}

\newcommand{\dya}[1]{|#1\rangle\langle#1|}

\newcommand{\encs}{\end{cases}}
\newcommand{\enm}{\end{matrix}}

\newcommand{\rB}{\textbf{r}}

\def\outl#1{\par{\medskip\noindent\hspace*{0.1cm}\bf
      \mathversion{bold}#1\mathversion{normal}\smallskip} }
   \def\xa{} \def\xb{}  

 \def\outl#1{}\def\xa{}\def\xb{}

\ca
 \def\outl#1{\par{\medskip\noindent\hspace*{.5cm}\bf
      \mathversion{bold}#1\mathversion{normal}\smallskip} }
 \long\def\xa#1\xb{} 
  
\cb

\begin{document}

\title{Reply to Comment on ``Multitime Quantum Communication: Interesting But
  Not Counterfactual'' by L. Vaidman}
\author{Robert B. Griffiths\thanks{Electronic address: rgrif@cmu.edu}\\
Department of Physics\\
Carnegie Mellon University\\
Pittsburgh, PA 15213}

\date{Version of 22 September 2023}
\maketitle

\begin{abstract}
  This is a response to comments and criticisms found in the preceding Comment
\cite{Vdmn23} by Vaidman on the paper \cite{Grff23}.
\end{abstract}




A significant part of Vaidman's Comment \cite{Vdmn23} is devoted to a
discussion of \emph{counterfactuals}, starting with a quotation from Penrose.
The use of counterfactuals in discussions of quantum foundations in fact goes
back much earlier, see e.g. \cite{Stpp71}. The basic idea involved in a
counterfactual is a comparison between two (or more) situations: one the
``actual'' world and the other the ``counterfactual'' world which differs from
the former in certain specified ways. One then considers various consequences
of these differences. In quantum theory this can lead to difficulties and
paradoxes when the physical properties of interest in the two worlds are
represented by incompatible observables or noncommuting projectors. The
consistent histories (CH) approach to quantum theory avoids such paradoxes by
refusing to compare incompatibles; see \cite{Grff99} and Ch.~19 of
\cite{Grff02c}. For the way in which CH resolves the (supposedly)
interaction-free measurement paradox mentioned by Vaidman, see Ch.~21 of
\cite{Grff02c}.

In \cite{Grff23}, the paper addressed by Vaidman's Comment, it is argued that
the claim of counterfactual communication by Salih et al.\ in \cite{Slao13}
fails in that it incorrectly assigns a probability for a photon to be in the
communication channel connecting Alice and Bob at intermediate times when
quantum interference effects are important, as well as incorrectly counting the
number of times it passes through the channel. While both errors are
significant, the first is more interesting in that it raises the question of
what can properly be said about a quantum particle's location at an
intermediate time given a wavefunction evolving unitarily from an initial state
on its way to a later measurement.

In Hilbert space quantum theory---which is to say the basic principles set
forth by von Neumann \cite{vNmn55b}, see in particular Sec.~III.5---a physical
property is represented by a projector $P$ on a Hilbert subspace. In particular
if the property is that the particle is in some spatial region $R$, the
projector $P$ applied to the position wavefunction $\psi(\rB)$ leaves it
unchanged for all $\rB\in R$, but sets it equal to zero elsewhere.
Consequently, if $\psi(\rB)$ is nonzero both for $\rB$ in some region inside
$R$ and also some other region outside $R$, the projector $\dya{\psi}$
corresponding to the particle's wavefunction (assumed normalized) does not
commute with $P$, and when projectors do not commute---this is the essence of
quantum uncertainty principles---there is no meaningful way to discuss whether
or not the particle is in $R$. A well-known example is the double slit paradox
where, in the presence of interference, one cannot meaningfully say which slit
the photon passed through. For this reason the CH interpretation of quantum
mechanics considers the conjunction of two properties represented by
noncommuting projectors to be meaningless: To say the particle is in or outside
$R$ makes no more sense than to discuss whether $S_x$ is $+1/2$ of $-1/2\hbar$
for a spin half particle when $S_z$ is $+1/2\hbar$.

Vaidman tries to get around this difficulty by asking whether a quantum
particle leaves a trace of its presence at a particular location via a weak
interaction with some other physical system at this location. That such a weak
measurement does not resolve the problem but simply generates more paradoxes
was known to Feynman; see his discussion in \cite{FylS651} of a weak light
source following the double slit---in his case the two holes with a coherent
electron wave passing through them. For an analysis of this situation based on
consistent quantum principles, see Sec.~13.5 of \cite{Grff02c}. Vaidman's
nested Mach-Zehnder paradox \cite{Vdmn13} has the same general character. Its
resolution when weak measurements are analyzed using consistent quantum
principles, \cite{Grff16}, was not discussed in Vaidman's Comment
\cite{Vdmn17} on that paper. The objections to the nested Mach-Zehnder paradox
by Englert et al.\ in \cite{Enao17} are similar: they argue that one cannot assign
a probability to a particle's following a particular path when it is in a
coherent superposition of amplitudes on different paths.

Towards the end of \cite{Vdmn23} Vaidman discusses the use of a quantity called
\emph{Cost}, used in \cite{Grff23} as a measure of channel usage. In response
it may be noted that Cost was introduced as a replacement for the misleading
use of ``probability'' in \cite{Slao13}, as in much of the succeeding
literature. In its favor is the fact that Cost is a well-defined mathematical
quantity in situations where probabilities cannot be consistently assigned, and
its use leads to the rigorous bound in Sec.~III~D of \cite{Grff23}, probably
the most interesting technical result in that paper. However, as with any novel
idea, only the future will show whether Cost is really useful or needs to be
replaced by something else. Vaidman's concern that the analysis using Cost
includes both cases in which a communication protocol succeeds as well as when
it fails is not relevant to the situation considered in \cite{Slao13}, where
the protocol always succeeds with high probability---an instance of what is
called a \emph{full} protocol in Sec.~III~C of \cite{Grff23}. The contrasting
case of \emph{partial} protocols---as, for example, when by convention the
non-arrival of a photon in Alice's apparatus at a particular time signals that
Bob has transmitted the bit 1---requires a separate discussion, which might be
a useful subject for some future paper.

I am grateful to Carnegie-Mellon University and its Physics Department for
continuing support of my activities as an emeritus faculty member.


\begin{thebibliography}{10}

\bibitem{Vdmn23}
Lev Vaidman.
\newblock Comment on ``Multitime quantum communication: Interesting but not
  counterfactual''.
\newblock arXiv:2306.16756, 2023.

\bibitem{Grff23}
Robert~B. Griffiths.
\newblock Multitime quantum communication: Interesting but not counterfactual.
\newblock {\em Phys. Rev. A}, 107:062219, 2023.
\newblock arXiv:2301.01730 v3.

\bibitem{Stpp71}
Henry~Pierce Stapp.
\newblock S-matrix interpretation of quantum theory.
\newblock {\em Phys. Rev. D}, 3:1303--1320, 1971.

\bibitem{Grff99}
Robert~B. Griffiths.
\newblock Consistent quantum counterfactuals.
\newblock {\em Phys. Rev. A}, 60:5--9, 1999.

\bibitem{Grff02c}
Robert~B. Griffiths.
\newblock {\em Consistent Quantum Theory}.
\newblock Cambridge University Press, Cambridge, U.K., 2002.
\newblock http://quantum.phys.cmu.edu/CQT/.

\bibitem{Slao13}
Hatim Salih, Zheng-Hong Li, M.~Al-Amri, and M.~Suhail Zubairy.
\newblock Protocol for direct counterfactual quantum communication.
\newblock {\em Phys. Rev. Lett.}, 110:170502, 2013.
\newblock arXiv:1206.2042.

\bibitem{vNmn55b}
John von Neumann.
\newblock {\em Mathematical Foundations of Quantum Mechanics}.
\newblock Princeton University Press, Princeton, 1955,2018.
\newblock Translation by R. T. Beyer of \textit{Mathematische Grundlagen der
  Quantenmechanik}, Springer-Verlag, Berlin, 1932.

\bibitem{FylS651}
R.~P. Feynman, R.~B. Leighton, and M.~Sands.
\newblock {\em The Feynman Lectures on Physics}, volume III: Quantum Mechanics.
\newblock Addison-Wesley, Reading, Mass., 1965.
\newblock Ch.~1.

\bibitem{Vdmn13}
L.~Vaidman.
\newblock Past of a quantum particle.
\newblock {\em Phys. Rev. A}, 87:052104, 2013.

\bibitem{Grff16}
Robert~B. Griffiths.
\newblock Particle path through a nested {M}ach-{Z}ehnder interferometer.
\newblock {\em Phys. Rev. A}, 94:032115, 2016.
\newblock arXiv:1604.04596.

\bibitem{Vdmn17}
L.~Vaidman.
\newblock Comment on `particle path through a nested mach-zehnder
  interferometer'.
\newblock {\em Phys. Rev. A}, 95:066101, 2017.
\newblock arXiv:1610.07734.

\bibitem{Enao17}
Berthold-Georg Englert, Kelvin Horia, Jibo Dai, Yink~Loong Len, and Hui~Khoon
  Ng.
\newblock Past of a quantum particle revisited.
\newblock {\em Phys. Rev. A}, 96:022126, 2017.
\newblock arXiv:1704.03722 v2;
\newblock U. Peleg and L. Vaidman
\newblock Comment on 'Past of a quantum particle revisited'
\newblock {\em Phys. Rev. A}, 99:026103, 2017.
\newblock  arXiv:1805.12171.
\end{thebibliography}

\end{document}